\documentclass[aps,prx,a4paper,showpacs,amsmath,amssymb,superscriptaddress,floatfix,footinbib,twocolumn]{revtex4-2}

\usepackage{hyperref}
\hypersetup{colorlinks,linkcolor=blue,urlcolor=blue,citecolor=blue}

\usepackage{bbm}

\usepackage[normalem]{ulem}
\usepackage{tabularx}
\usepackage{graphics,graphicx}
\usepackage{epsfig}

\usepackage{xcolor}
\usepackage{float}
\usepackage{adjustbox}

\usepackage{dsfont}

\usepackage{amsmath,amssymb,mathrsfs}
\usepackage{qcircuit}
\usepackage{verbatim}

\usepackage{graphicx} 

\begin{document}

\title{Automated Discovery of Gadgets in Quantum Circuits for \\
       Efficient Reinforcement Learning
      }


%
%

\author{Oleg M. Yevtushenko}
\affiliation{Max Planck Institute for the Science of Light, Erlangen, Germany}
\affiliation{Friedrich-Alexander-Universit{\"a}t Erlangen-N{\"u}rnberg, Germany}

\author{Florian Marquardt}
\affiliation{Max Planck Institute for the Science of Light, Erlangen, Germany}
\affiliation{Friedrich-Alexander-Universit{\"a}t Erlangen-N{\"u}rnberg, Germany}

\begin{abstract}
Reinforcement learning (RL) has proven itself as a powerful tool for the discovery of quantum circuits and quantum protocols. We have recently shown that including composite quantum gates -- referred to as ``gadgets'' -- in the action space of RL agents substantially enhances the RL performance \cite{olle_2025} in the context of quantum error correction. However, up to now the gadgets themselves had to be constructed manually. In this paper, we suggest an algorithm for the automated discovery of new gadgets and families of related gadgets. The algorithm is based on the representation of quantum circuits as directed graphs and an automated search for repeated subgraphs. The latter are identified as gadgets. We demonstrate the efficiency of the algorithm, which allows us to find two new gadget families suitable for RL. We compare the performance of 4-qubit gadgets taken from a previously known and a newly discovered family and discuss their advantages and disadvantages.
\end{abstract}

\date{\today}

\maketitle

Reinforcement learning has emerged as a powerful tool for automatically discovering improved strategies in the control of quantum computers (starting with the early works \cite{fosel2018reinforcement,niu2019universal,Nautrup_2019,andreasson2019quantum}; for a recent review see \cite{krenn2023artificial}). In particular, this includes the automated discovery of quantum error correction (QEC) strategies \cite{fosel2018reinforcement,Nautrup_2019,andreasson2019quantum,sweke2020reinforcement,olle_2024,zen2024quantum} as well as tasks like quantum circuit optimization \cite{foesel2021quantumcircuitoptimizationdeep,ruiz2024quantumcircuitoptimizationalphatensor}. All of these examples deal with quantum circuits consisting of elementary quantum gates.

We have recently demonstrated \cite{olle_2025} that the power of the automated discovery of quantum circuits via reinforcement learning (RL) is crucially enhanced if one expands the action space of the RL agents by adding {\it hierarchical families of gadgets} to the elementary one- and two-qubit gates. In the concrete case considered in  \cite{olle_2025}, the task was to find quantum circuits for logical qubit encoding in the context of QEC. In that scenario, each gadget is a certain combination of CNOTs, which entangles several qubits. A gadget family is based on a hierarchy of generations: each child generation can be produced as a combination of gadgets from the parent generation.

The gadget-based RL (gRL) introduced in  \cite{olle_2025} significantly outperforms the speed and the success rate of the standard RL approach. We have demonstrated this in the example of the discovery of QEC codes \cite{roffe_2019}, see Refs.\cite{olle_2024,olle_2025}. The enhanced performance of the gRL has allowed us to discover for the first time encoders of the codes $ [[n, 1, d]] $, with $ d = 6, 7$, and $ [[36,7,6]] $. This was a noticeable breakthrough in the automated discovery of quantum circuits, at scales which correspond to the current state-of-the-art QEC hardware \cite{krinner2022realizing,ryan2021realization,postler2022demonstration,cong2022hardware,GoogleQuantum2023,sivak2023real,google2024quantum}.

In Ref.\cite{olle_2025} , we identified gadgets by visual analysis of simple RL-generated circuits. This human-in-the-loop procedure was a noticeable disadvantage of our approach. For example, we did not know whether the hierarchical family of gadgets discovered in Ref.\cite{olle_2025} was the only one suitable for RL. Automating the search of gadget families would provide a solution to this challenge. If other families exist one can choose the family which seems most appropriate for certain goals. This would pave the way for further optimization of quantum circuit discovery by RL agents.

\begin{figure}
    \centering
    \includegraphics[width=0.75\columnwidth]{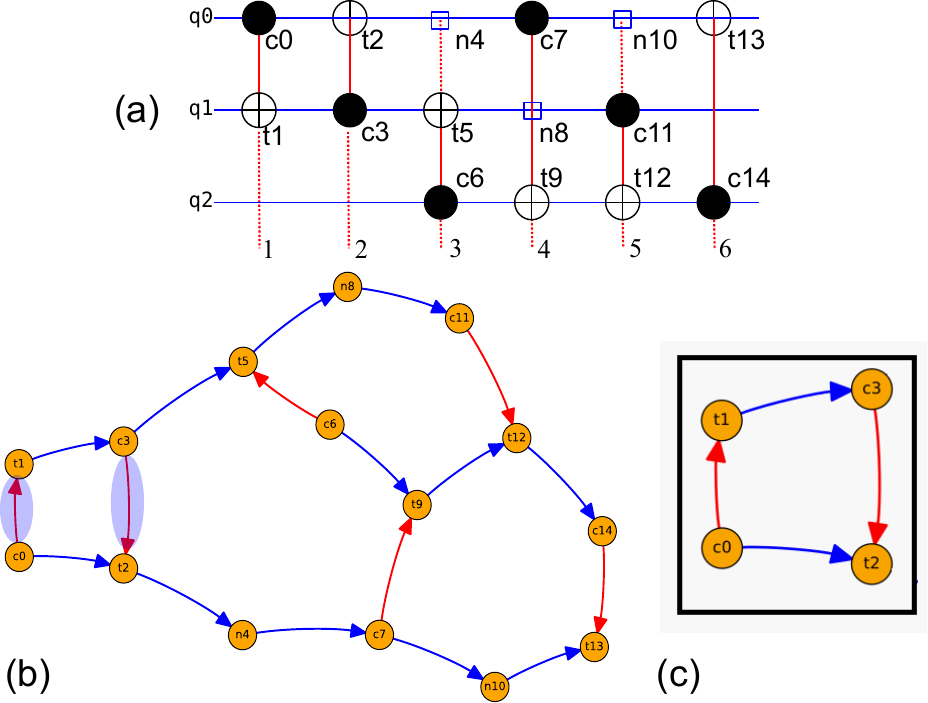}
    \caption{\textbf{Converting a quantum circuit to a graph.} (a) The initial circuit (red solid lines: CNOTs, blue lines: time evolution); (b) its graph representation where unnecessary empty, i.e. with CNOT nodes, external temporal lines  have been excluded (red arrows: CNOTs, blue arrows: temporal links); (c) a suitable subgraph, which has been obtained after selecting the two  CNOTs highlighted in panel (b) -- see explanations in the text \cite{sm-1}. Dotted extensions of the red lines in Panel (a) indicate the temporal layers of the circuit. To make the transformations more transparent, we denote control ("c"), target ("t"), and neutral ("n") nodes by unique labels in all panels.}
    \label{fig:converting}
\end{figure}

\begin{figure}[t]
    \centering
    \includegraphics[width=0.9\columnwidth]{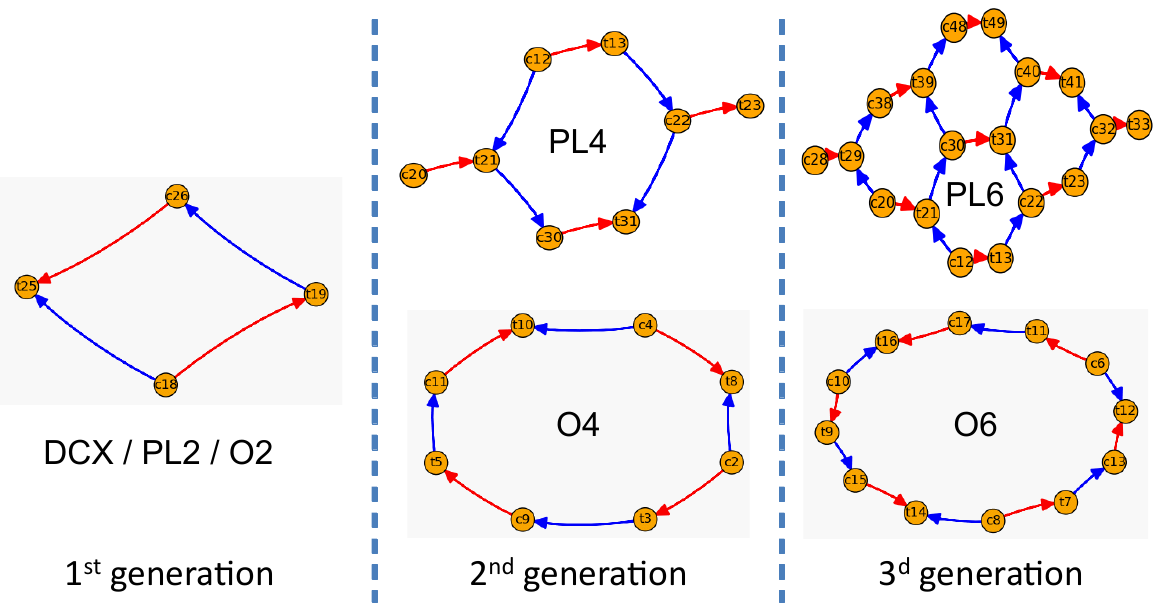}
    \caption{\textbf{Two new families of gadgets automatically discovered by the algorithm.} The PL- and O-families have been discovered by applying the algorithm described in the main text to the encoding circuits with simple to medium complexity. These two families
    are best detectable in the codes [[9,1,3]] and [[11,1,3]], respectively. Examples of circuits where the corresponding gadgets have been found, are shown in the Suppl.Mat.\ref{sm:gadget_generation}. Integer numbers after the letters "PL" and "O" denote the number of qubits entangled by a given gadget. Subsequent generations can be constructed based on the structure of the generations 1-3.}
    \label{fig:families}
\end{figure}

\begin{figure}
    \centering
    \includegraphics[width=0.6\columnwidth]{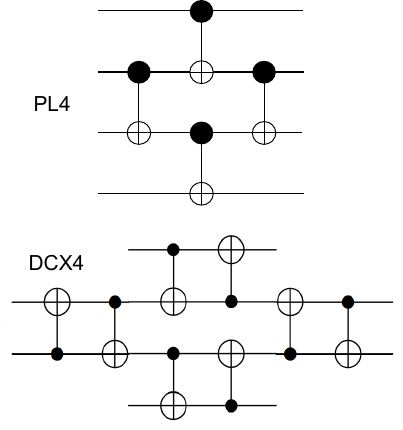}
    \caption{\textbf{4-qubit gadgets from the PL- and DCX families.} The performance of these gadgets is compared in Fig.\ref{fig:performance}
            }
    \label{fig:4Q_gadgets}
\end{figure}

\begin{figure}
    \centering
    \includegraphics[width=0.9\columnwidth]{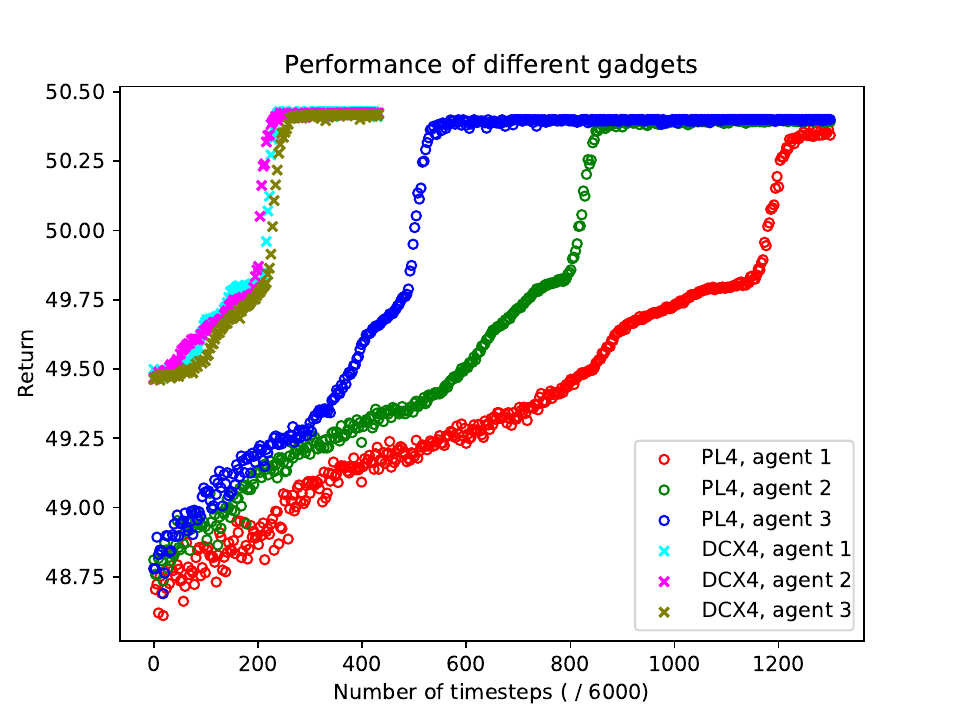}
    \caption{\textbf{Performance of gadgets from different families.} The speed of convergence is shown for discovering encoding circuits of the [[29,1,6]] code. The RL procedure can find solutions when the return approaches values above 50 and saturates. In this example,  the speed of (CX+DCX4)-based calculations is from 2 to 6 times higher than that of (CX+PL4)-based ones. Our attempts to discover these circuits by using only CX gates failed. For clarity, only every third point is shown, and the data obtained by using the DCX4 gadgets is cut when all curves saturate at a plateau.}
    \label{fig:performance}
\end{figure}

{\it In the current paper}, we remove the above-mentioned restriction of the approach of Ref.\cite{olle_2025}. Namely, we 1) introduce an algorithm, which can be used for automating the discovery of gadget families (see Fig.\ref{fig:converting}), 2) showcase two new families of gadgets, which have been found with the help of the suggested algorithm (see Fig.\ref{fig:families}), and 3) compare the performance of two representatives taken from the old and the newly discovered gadget families (see Figs.\ref{fig:4Q_gadgets},\ref{fig:performance}).

A systematic study of the new gadget's performance has led us to the following conclusions. Firstly, the new, automatically discovered gadget families seem to be appropriate for RL, boosting the performance under certain conditions. Secondly, since each family has advantages and disadvantages, the choice of the family for the gRL calculations should  correspond to the objectives pursued.

Let us start with outlining our {\it algorithm for the automated discovery of gadgets in quantum circuits}, which we have developed and successfully used. Our approach
is based on a representation of a given quantum circuit as a directed graph. The first basic steps of the algorithm are illustrated in Fig.\ref{fig:converting}. In that example, we have used a circuit consisting of three qubits and six CNOT gates.

To convert this circuit into a graph, we lay out the circuit in the form of a space-time grid, where "space" corresponds to the different qubits and all the grid points will become nodes of the graph. As usual in quantum circuits, horizontal blue lines, drawn for each qubit, denote the time evolution. To form the grid, we also introduce regularly spaced temporal layers, which are shown in panel (a) as vertical red lines. Red solid vertical lines represent CNOT gates while red dotted lines are added to highlight the layers. We label all the nodes of the graph in the following way: We denote the control or target node of any CNOT as "c" or "t", respectively, while "n" marks any other node. In addition, each node is assigned a unique number. Edges corresponding to CNOTs will be drawn as red arrows when depicting the graph (pointing \textbf{from control to target}), while edges corresponding to time evolution will be drawn as blue arrows (pointing into the direction of increasing time). We note that any time-like edges at the beginning and at the end of the circuit that do not contain CNOT nodes are unimportant for gadget discovery and are, therefore, neglected. After these steps, we obtain the directed graph shown in Fig.\ref{fig:converting}(b).

Based on our previous experience, we surmise that the gadgets correspond to closed parts of the graph. Therefore, we delete path- and star-like parts of the graph at all stages of the gadget discovery \cite{sm-1}.

When trying to discover any gadget containing exactly $ C_{f} $ CNOT gates, we proceed in the following way: We  choose from the prepared graph all possible nonequivalent combinations containing exactly $C_g$ CNOT gates (edges denoted by red arrows) and all time-lines (edges denoted by blue arrows) connecting these gates.
This procedure yields $ C_{T}! / C_{f}! (C_{T}-C_{f})! $ subgraphs, with $ C_{T} $ being the total number of CNOTs in the analyzed circuit. The subgraph is suitable for further analysis if it is connected and closed.
An example with $ C_g = 2 $ is shown in Fig.\ref{fig:converting}(c) \cite{sm-1}.

To apply this approach for discovering gadgets in quantum circuits used for QEC logical qubit encoding, we proceed like the following: We  choose a simple code $ [[n,k,l]] $, generate its encoders with the help of our standard RL approach, select the encoders with different canonical tableaux, and apply the above described procedure to each such encoder. In this way, we generate a database of subgraphs from which we will try to identify the gadgets. To make the search of gadgets more constructive, we apply several additional constraints \cite{sm-1}. Firstly, the selection procedure illustrated in Fig.\ref{fig:converting} may accidentally cut out some CNOT line keeping one of its end nodes. This would lead to the appearance of ``empty'', i.e. not connected to any red arrow, c- or t- node(s). We believe that the subgraphs containing such empty nodes are not appropriate for the automated gadgets discovery and exclude them from further analysis. Second, all selected subgraphs must be ``stationary", in the sense that the primitive gates (CNOTs) do not commute. The nonstationary subgraphs (where many orderings would be possible) may possess ambiguity, and we also exclude them.

We would like to mention that the suggested procedure of preprocessing the graph and finding suitable subgraphs is rather restrictive and based on our previous experience with gadget discovery by manual inspection. We think the procedure may be improved in the future by developing a corresponding more rigorous mathematical background.

Finally, we check whether the resulting database of appropriate subgraphs contains subgraphs which are isomorphic, in which case we treat them as equivalent. For each subgraph, we count the number of its repetitions (including all equivalent counterparts) in the database, $ N_{r} $. If a given subgraph has $ N_r > 1 $ (or larger than a cutoff value, $ N_r > N_c $) we identify it as a gadget. This concludes our algorithm for automated gadget discovery.

We emphasize that $ N_r $ reflects the number of repetitions of the subgraph in the entire collection of quantum circuits. We do not request the repetitions to occur in the same circuit. For all standard graph operations, we have used procedures from the python library ``igraph''.

In our previous work, we had used visual analysis of nonequivalent encoders \cite{olle_2025} to identify repeated circuit elements, i.e. gadgets. This simplest approach had allowed us to identify only one family, which starts from the double-CX gadget with two CNOTs, DCX in the notation of Ref.\cite{olle_2025}, and has all family members consisting of the DCX elementary blocks. To reflect the geometry of such gadgets, we refer here to this family as ``the DCX-family''. The DCX gadget entangles 2 qubits. An example of more complicated gadget from the DCX-family, which entangles 4 qubits, is shown in Fig.\ref{fig:4Q_gadgets}.

To check the existence of other gadget families, we have applied the suggested algorithm to encoders of QEC codes with $ n \ge 7 $, $ k = 1 $, and $ d \ge 3 $. Our choice of the lower limits is governed by the necessity to have a large number of relatively long circuits where repeating patterns show up. On the other hand, encoders of too complicated codes appeared to be inefficient for gadget discovery. This might be related to purely combinatorial reasons, namely, the number of encoders grows exponentially with increasing $ n $ and $ d $, and we were unable to analyze a large enough fraction of the circuits where the repeating subgraphs would be noticeable. Therefore, we restricted ourselves to the automated analysis of codes with simple to medium complexity, $ 7 \le n \le 15 $, $ d = 3, 4$.

Our approach has immediately revealed two {\it new families of gadgets}, which we call ``the PL-family'' (for gadgets which resemble stylized "+"-signs) and "the O-family" (for gadgets which resemble "O"-letters), see the first 3 generations of these families shown in Fig.\ref{fig:families}. We note that the gadgets look quite different in the graph representation and in the original circuits where they were found, see Suppl.Mat.\ref{sm:gadget_generation} for details.

Finally, we have checked that the newly discovered gadgets are suitable for gRL. We have done this by running the RL agents for the same codes as before but adding the gadgets of the second generation, either PL4 or DCX4 (see Fig.\ref{fig:4Q_gadgets}) to the RL action space \cite{DCX_not}. Note that these are simplest gadgets which are different in the DCX- and PL families, cf. Fig.\ref{fig:families}.

\begin{figure}
    \centering
    \includegraphics[width=0.9\columnwidth]{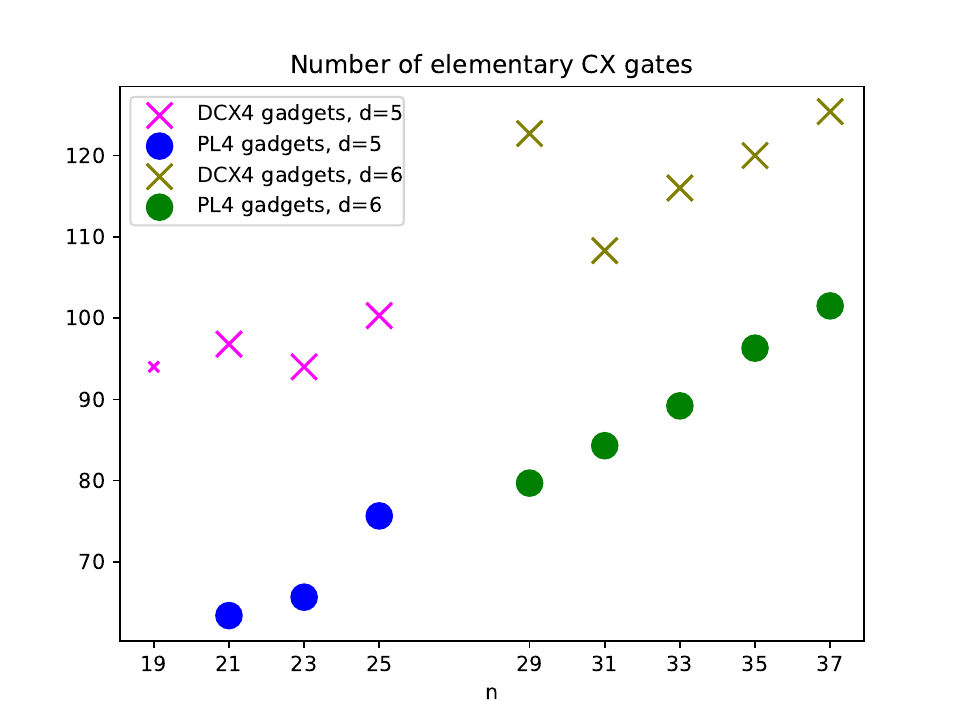}
    \caption{\textbf{Average number of elementary CX gates needed for solutions found with the help of gadgets from different families.}
    Large symbols correspond to a success rate of $\sim$100\%. The small x-symbol reflects the success rate of the d4cx-based RL, about 30\% for the code [[19,1,5]]. This is a more challenging case where the PL4-based RL failed. After increasing the connectivity, the RL procedure finds answers for the code [[19,1,5]] at almost the same success rate for both gadget types.
           }
    \label{fig:CX_length}
\end{figure}

Two important indicators that we have investigated are the speed of the RL search for the solutions and its success rate. The latter has been defined as the fraction of training runs which arrive at the desired solution, i.e. a circuit producing the desired code. We have addressed the encoders of the code [[29,1,6]] to explore the speed, see Fig.\ref{fig:performance}.
The qubit connectivity has been chosen as next-to-nearest neighbors.
Note that we were unable to find these encoders with the help of the standard RL procedure \cite{olle_2024}, due to their complexity. Both gadgets, DCX4 or PL4, enable gRL to find the solutions. We have observed that the RL training finds solutions faster if the agent is allowed to use the combination "CX + DCX4". Simultaneously, this combination provides a higher success rate for the codes which are most complicated for the RL approach, see the result for the code [[19,1,5]] in Fig.\ref{fig:CX_length} and its discussion in the caption of that figure. These are potential advantages of the DCX-family. As argued below,
we attribute them to the larger number of elementary CX gates per one gadget of the DCX family (8 per DCX4 and only 4 per PL4).

The downside of the larger size of the previously employed DCX4 gadget as compared to the newly discovered PL4 reveals itself in the substantially larger number of elementary CNOT gates in those circuits which are found by relying on the "CX + DCX4" combination, see Fig.\ref{fig:CX_length}. The results shown in this figure have been averaged over several random initializations of neural networks. This finding can also be explained by the larger number of CNOTs entering DCX4.
The average number of PL4 gadgets per circuit of the given code $[[n,k,d]]$ is larger (which explains a bit the reduced speed of the RL search), but the average number of elementary gates is smaller.  Hence, the smaller average number of DCX4 gadgets cannot compensate the larger number of the CNOT gates included within this gadget.

Both combinations, "CX + DCX4" and "CX + PL4", yield relatively large weights of generators, see Suppl.Mat.\ref{sm:weights}, which are almost the same for the gadgets from the different families and slightly increase with increasing the number of qubits $ n $. The large weights can be suppressed to some extent if the RL reward is biased towards solutions with smaller weights.


{\it Conclusions/discussion}:
We have suggested an algorithm which allows one to perform an automated search for gadgets (building blocks) suitable for the accelerated RL-based discovery of large-scale quantum circuits. The basic gadget involves a minimal number of entangled qubits, 2 in our study. More complicated gadgets act on 4, 6, \ldots qubits. Gadgets with a similar structure are called a gadget family. The number of entangled qubits defines a generation. Inside the family, the gadgets of a given child's generation often can be constructed as a combination of the gadgets from the previous parent generation. Such families therefore have a hierarchical structure.

The proposed algorithm is based on the possibility of representing the quantum circuits as graphs. Following Ref.\cite{olle_2025}, we have restricted ourselves to the application of gRL to the case of quantum error correction. To find the gadgets, one has to generate encoders of a given code with the help of the standard (not gadget-based) RL.
Next, one processes the respective quantum circuits and identifies subgraphs which obey certain selection rules.
This procedure yields a database of the subgraphs. Repeated subgraphs from this database, including their isomorphic counterparts, are identified as the gadgets.
%
The automated approach presented in this paper has allowed us to find two new families of gadgets.

To check whether the newly found families are suitable for gRL, we have compared the performance of two 4-qubit gadgets, DCX4 from Ref.\cite{olle_2025} and the new gadget PL4. Our study shows that both gadgets substantially accelerate the search for solutions in gRL. Using the DCX4, the gRL agents could find solutions faster with a higher success rate in complicated setups. However, the DCX4-based circuits contain a noticeably larger number of elementary CNOT gates.

Overall, having available multiple gadget families discovered by our approach allows one to choose an optimal family for gRL depending on certain criteria. In our example, we believe that DCX$j$ gadgets are more efficient for finding long and complicated encoders while PL$j$ gadgets allow one to minimize the number of elementary gates per one encoder.

 We note in passing that neither our understanding of gadgets nor of the rules for processing graphs and subgraphs are based on a mathematically rigorous analysis. Rather, in this paper, we have presented a summary of our successful numerical experiments. The development of the mathematical basis, which is far beyond the scope of the present paper, would  make the search for new gadgets more efficient.

\acknowledgements

We gratefully acknowledge the participation of Jan Oll{\'e} Aguilera at the early stages of this work and helpful discussions
with Remmy Zen.
This research is part of the Munich Quantum Valley network, which is supported by the Bavarian state government with funds from the Hightech Agenda Bayern Plus.

\bibliography{refs}

\begin{thebibliography}{21}
\bibitem [{sm-()}]{sm-1}%
  \BibitemOpen
  \href@noop {} {}\bibinfo {note} {More details can be found in
  Suppl.Mat.\ref{sm:pre_pross}}\BibitemShut {NoStop}%
\end{thebibliography}

\appendix

\begin{widetext}

\section{Preprocessing circuits and selecting appropriate subgraphs}
\label{sm:pre_pross}

In the main text, we have briefly explained the main steps which are needed for the automated discovery of gadgets, see also Fig.\ref{fig:converting}. They include converting the quantum circuits into the graph representation, pre-processing it,
generating connected closed subgraphs with a given number of elementary CNOT actions, and selecting the subgraphs which are suitable for the automated search of repeated patterns. These patterns form the gadgets. Let us now elaborate the above steps in more detail, starting from the same example as that shown in Fig.\ref{fig:converting}.

\begin{figure}[h]
    \centering
    \includegraphics[width=0.8\columnwidth]{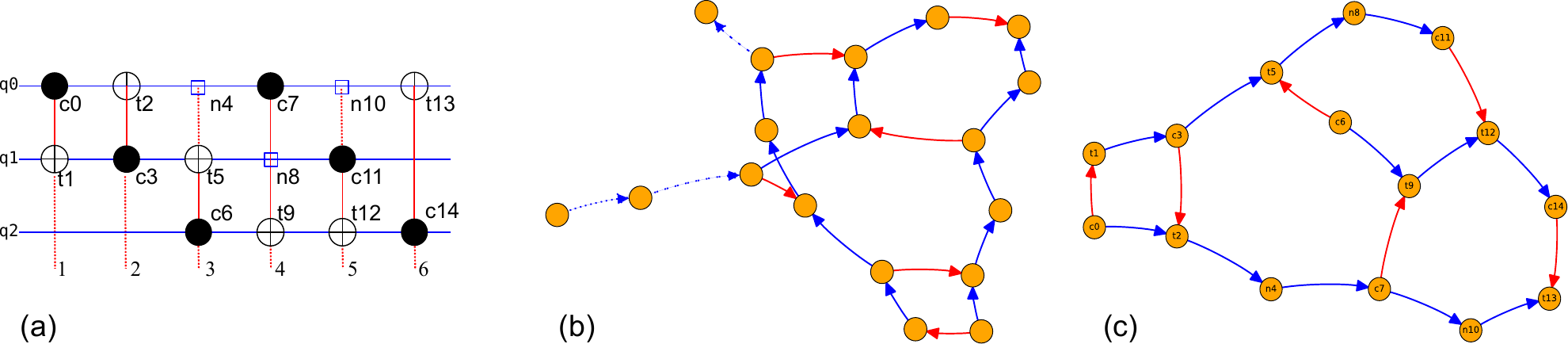}
    \caption{Converting the quantum circuit, panel (a), to a directed graph, panel (b), and deleting time-like edges, which result from the beginning or/and the end of the circuit and do not contain CNOT nodes.
    The final result of this procedure is shown in panel (c).}
\label{fig:converting_steps}
\end{figure}

The conversion of the quantum circuit to the directed graph is shown in Fig.\ref{fig:converting_steps}. Panel (b) of that figure illustrates in more detail the transition from panel (a) to panel (b) of Fig.\ref{fig:converting}.

%

Next, one has to generate subgraphs which are suitable for discovering gadgets. We assume that the selected $ C_g $ CNOT gates must belong to the connected and closed part of the graph. After their selection, the remaining $ C_T - C_g $ CNOT gates are neglected and time-like edges, which do not contain CNOTs, are neglected afterward. This produces the closed and connected subgraph. An example is shown in Fig.\ref{fig:selecting_steps} which illustrates the transition from panel (b) to panel (c) of Fig.\ref{fig:converting}.

\begin{figure}[t]
    \centering
    \includegraphics[width=0.7\columnwidth]{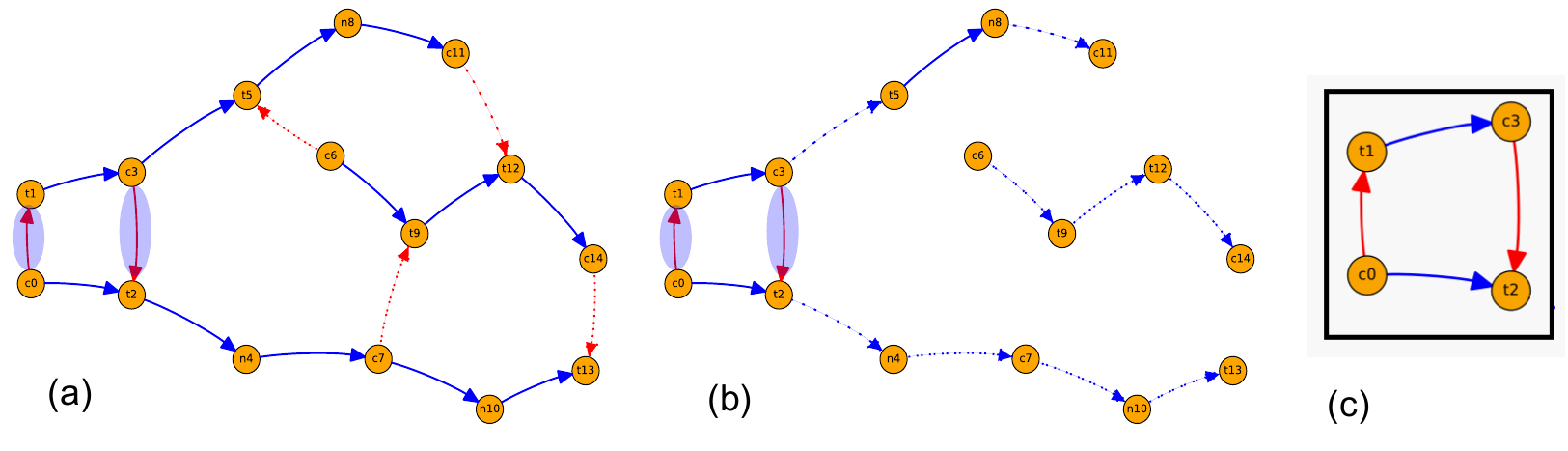}
    \caption{Selecting some CNOT-lines and deleting all other CNOT-lines, (a), deleting disconnected parts and empty lines which do not contain CNOT-lines, (b). The final result of this procedure is shown in panel (c).}
\label{fig:selecting_steps}
\end{figure}


 To make the search for gadgets more constructive, we apply several additional constraints. Firstly, the selection procedure illustrated in Fig.\ref{fig:selecting_steps} may erroneously cut out CNOT lines which are present in a larger gadget with a larger number of CNOTs. An example is shown in the left panel of Fig.\ref{fig:filtering} where one CNOT line has been cut out from the c-vertex, while this line enters a larger connected subgraph, see the right panel of Fig.\ref{fig:filtering}. We avoid such subgraphs with an "empty" c- or t- nodes and exclude them from further analysis.

 Second, all selected subgraphs must be stationary in the sense that primitive CNOT gates do not commute. The example shown in the right panel of Fig.\ref{fig:filtering} is stationary. We exclude nonstationary subgraphs. We note in passing that the stationarity of the subgraphs can be analyzed only in the initial graph which contains all relevant vertices. Such an advanced analysis is beyond the scope of the current paper and we restrict ourself to the check of stationarity of the automatically generated gadgets.

\begin{figure}[h]
    \centering
    \includegraphics[width=0.35\columnwidth]{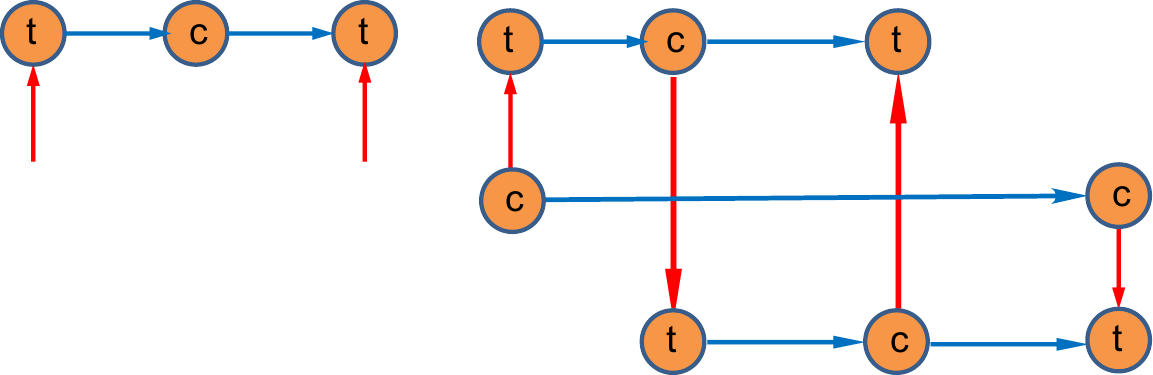}
    \caption{Left panel: An example of a sub-graph's part where one CNOT line has been cut out from the c-vertex, while this line enters a connected subgraph with a larger number of CNOT lines [right panel]. We exclude such subgraphs with an "empty" c- or t- vertex.}
\label{fig:filtering}
\end{figure}

\section{Gadgets of higher generations}
\label{sm:gadget_generation}

Fig.\ref{fig:families} shows how the gadgets of the PL- and O- families look in the graph representation. Their appearance in the quantum circuits may be quite different and not always easily recognizable by the naked eye. Fig.\ref{fig:cirC_gamilies} demonstrates several examples of the gadgets PL4, PL8, O4, and O6 found by our algorithm in relatively simple circuits. We have observed that the standard RL agents construct the gadgets more frequently when the number of qubits $ n $ is not large and the distance is small. Therefore, it suffices to use either simple or moderately complicated circuits to discover the gadget families.

\begin{figure}[h]
    \centering
    \includegraphics[width=0.4\columnwidth]{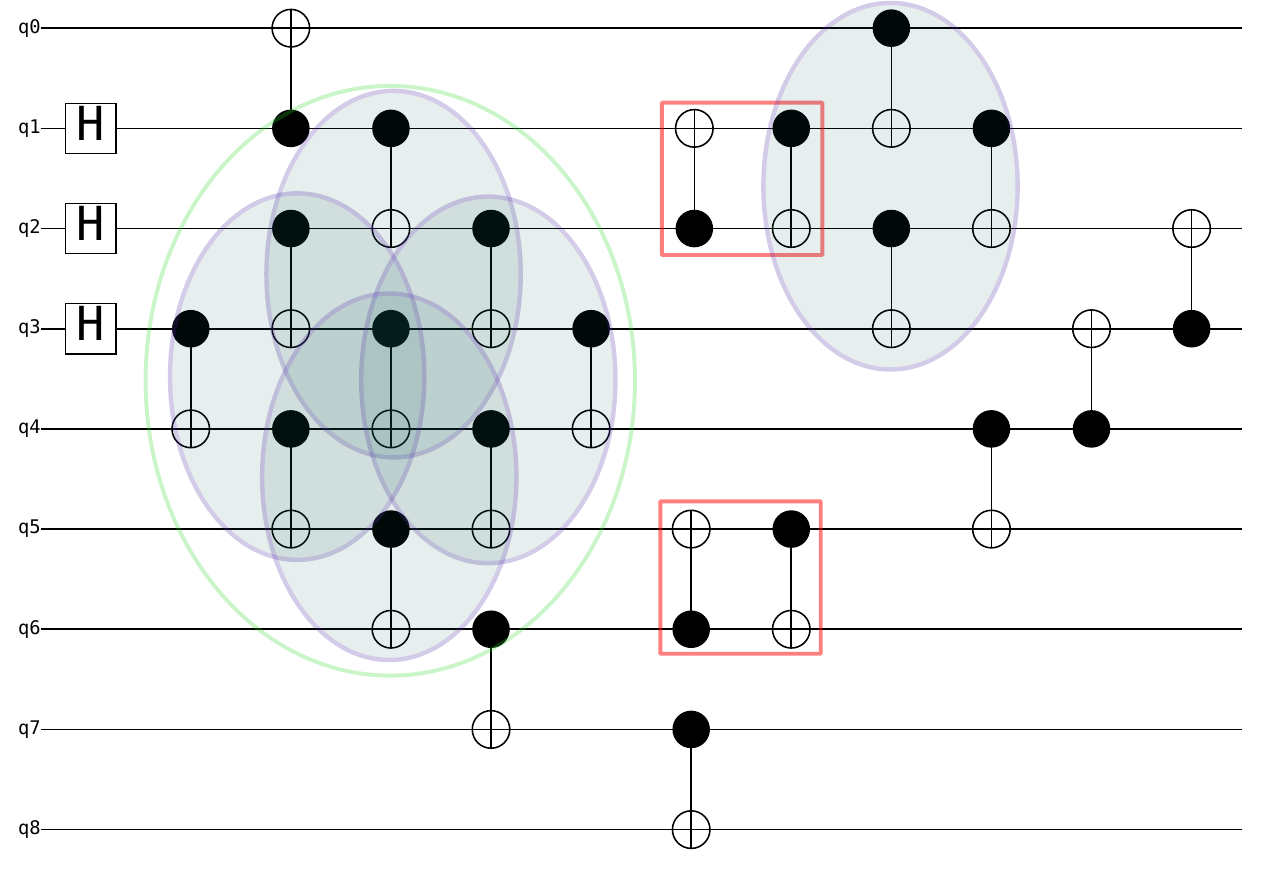}
    \hspace{1cm}
    \includegraphics[width=0.35\columnwidth]{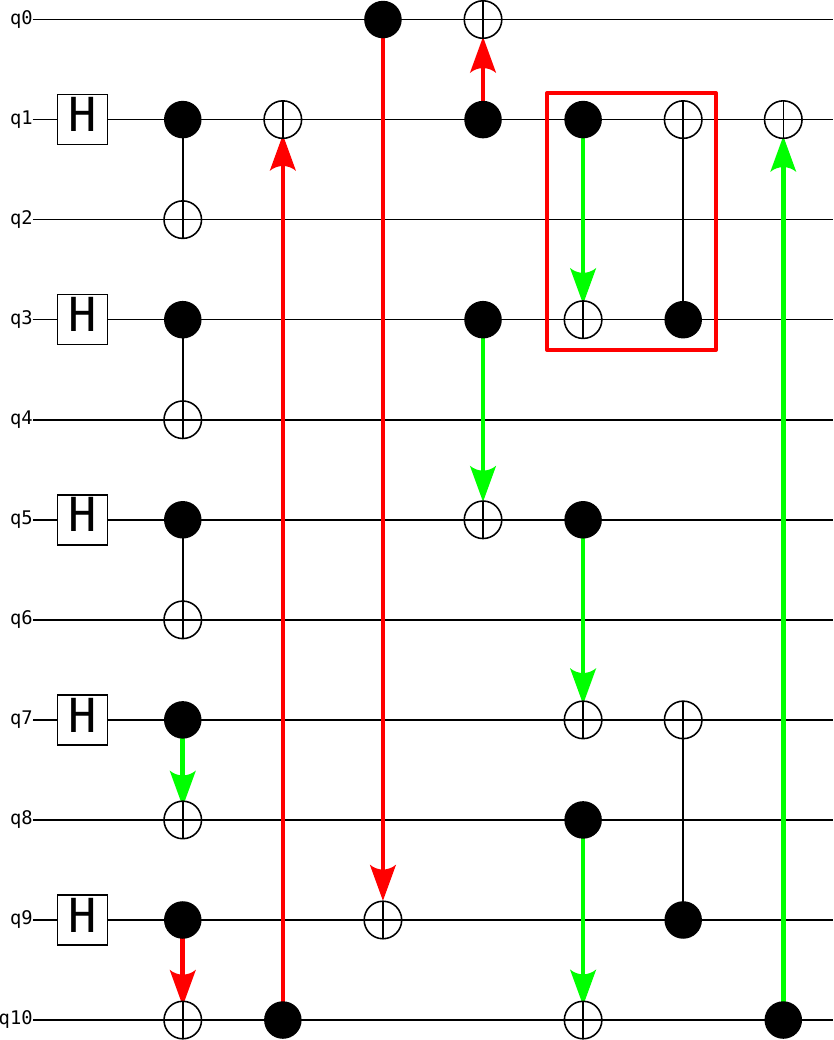}
    \caption{Examples of gadgets from the PL- and O-families which have been automatically found in simple circuits. Left panels: the gadgets from the PL-family which can be detected in the [[9,1,3]] codes. The circuit includes simultaneously  PL4 (shaded ovals) and PL6 gadgets (the region of nested PL4 gadgets). Right panel: the gadgets from the O-family which can be detected in the [[11,1,3]] codes. The gadgets O4 and O6 are highlighted by red and green lines of the elementary CNOT gates, respectively. Though the circuit may contain only one gadget of a given type, e.g. PL6 or O4 and O6, these gadgets are repeated in different encoders of the respective code. The basic gadgets containing two qubits only are highlighted by red squares in both panels.}
\label{fig:cirC_gamilies}
\end{figure}

\section{Weights of generators}
\label{sm:weights}

We have discussed in Ref.\cite{olle_2025} that using gadgets in the RL discovery of encoders increases the weights of the generators. The weights further increase with increasing complexity of the quantum circuits. These properties seem to be a common disadvantage of various families of gadgets and are only slightly different for the gadgets from the different families. An example of the n-dependence of the weights is shown in Fig.\ref{fig:weights_comp}. The encoders found with the help of the DCX4 and PL4 gadgets have, on average, rather similar generator weights.

\begin{figure}[t]
    \centering
    \includegraphics[width=0.5\columnwidth]{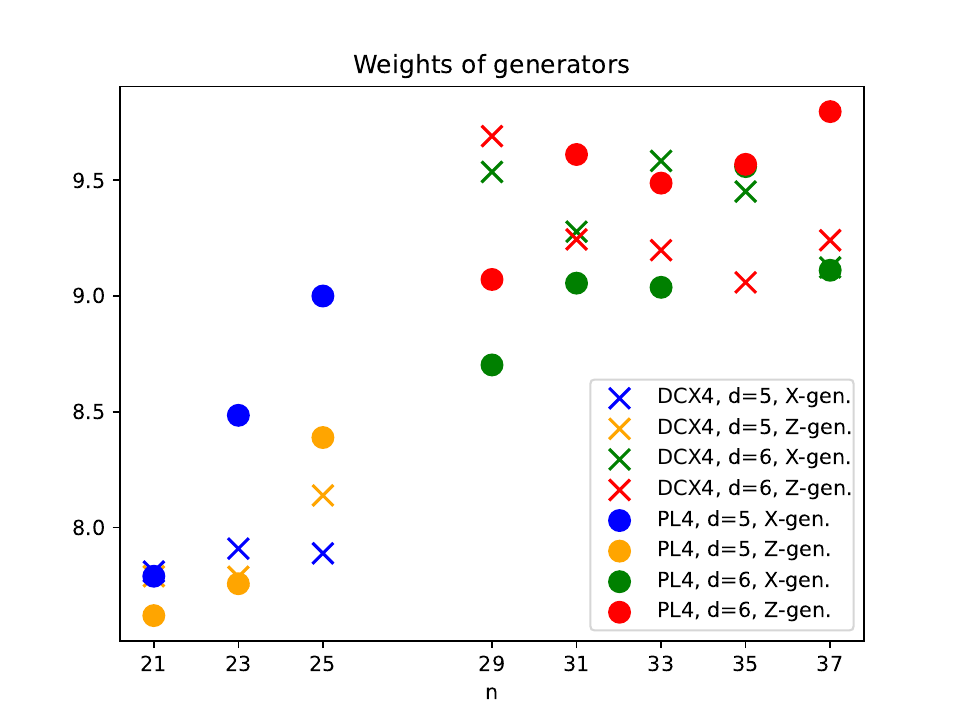}
    \caption{Average weights of codes [[n,1,5]] and [[n,1,6]]. Encoders have been discovered with the help of DCX4 (x-symbols) and PL4 (circles) gadgets for different values of the qubit number $ n $. The success rate in these calculations is always 100\%.
           }
    \label{fig:weights_comp}
\end{figure}

We note that the large weights inherent to encoders discovered by the gadget-based RL can be suppressed if one biases the reward function towards solutions with smaller weights of generators \cite{olle_2025}. This results in a trade-off between smaller weights and larger distances, which could complicate the search. However, the generator weights of the discovered encoders still remain higher than those of, for instance, surface codes.


\end{widetext}

\end{document}